\documentclass[twocolumn, times, trackchanges]{aastex63}

\usepackage{amsmath}
\usepackage{amssymb}
\usepackage{graphicx}
\usepackage{booktabs}
\usepackage{microtype}
\usepackage[all]{hypcap}
\usepackage[normalem]{ulem}
\usepackage{needspace}

\definecolor{linkcolor}{rgb}{0.0,0.3,0.5}

\begin{document}

\title{Milky Way Satellites Shining Bright in Gravitational Waves}
\shorttitle{MW Satellites Shining Bright in GWs}
\shortauthors{Roebber, Buscicchio, Vecchio et al.}

\newcommand{\bham}{School of Physics and Astronomy \& Institute for Gravitational Wave Astronomy, University of Birmingham, Birmingham, B15 2TT, UK}
\newcommand{\nrc}{National Research Council of Canada, Herzberg Astronomy \& Astrophysics Research Centre,\\ 5071 West Saanich Road, Victoria, BC V9E 2E7, Canada}
\newcommand{\aei}{Max-Planck-Institut f\"ur Gravitationsphysik (Albert-Einstein-Institut), D-30167 Hannover, Germany}

\author[0000-0002-5709-4840]{
 Elinore Roebber}
\affiliation{\bham}

\author[0000-0002-7387-6754]{Riccardo Buscicchio}
\affiliation{\bham}

\author[0000-0002-6254-1617]{Alberto Vecchio}
\affiliation{\bham}

\author[0000-0002-2527-0213]{Christopher J.\ Moore}
\affiliation{\bham}

\author[0000-0001-5438-9152]{Antoine Klein}
\affiliation{\bham}

\author[0000-0002-6725-5935]{Valeriya Korol}
\affiliation{\bham}

\author[0000-0002-2998-7940]{Silvia Toonen}
\affiliation{\bham}

\author[0000-0002-0933-3579]{Davide Gerosa}
\affiliation{\bham}

\author[0000-0003-4177-0098]{Janna Goldstein}
\affiliation{\bham}

\author[0000-0003-0259-858X]{Sebastian M.\ Gaebel}
\affiliation{\bham}\affiliation{\aei}

\author[0000-0003-1428-5775]{Tyrone E.\ Woods}
\affiliation{\bham}\affiliation{\nrc}

\correspondingauthor{Elinore Roebber}
\email{e.roebber@bham.ac.uk}

\date{\today}

\begin{abstract}
The population of Milky Way satellite galaxies is of great interest for cosmology, fundamental physics, and astrophysics.  They represent the faint end of the galaxy luminosity function, are the most dark-matter dominated objects in the local Universe, and contain the oldest and most metal-poor stellar populations. Recent surveys have revealed around 60 satellites, but this could represent less than half of the total.  Characterization of these systems remains a challenge due to their low luminosity.  We consider the gravitational wave observatory LISA as a potential tool for studying these satellites through observations of their short-period double white dwarf populations. LISA will observe the entire sky without selection effects due to dust extinction, complementing optical surveys, and could potentially discover massive satellites hidden behind the disk of the Galaxy.

\null
\end{abstract}

\section{Introduction}

The identification and characterization of  Milky Way (MW) satellite galaxies lie at the intersection of several outstanding problems in cosmology, astrophysics and fundamental physics \citep{BullockBoylan-Kolchin:2017}.  These include the nature of dark matter, the formation and evolution of the faintest galaxies, and their reionization history. Faint satellites also offer us the opportunity to study star formation in low-metallicity environments and systems chemically different compared to the MW, which may be relevant for the origin of r-process and heavy elements. 

Following the serendipitous discovery of the Sculptor dwarf galaxy~\citep{Shapley:1938}, only a dozen other MW satellites were known up until approximately 2010.  The Sloan Sky Digital Survey (SSDS) (and subsequently DECam, DES and Pan-STARRS, with the recent addition of Gaia \citep{Gaia}) has transformed the field raising the number to around 60; however, at least twice as many satellites are thought to exist, and the number could be nearly an order of magnitude higher \citep{Joshua:2019}. The observational effort to complete the census of the MW satellites is made particularly arduous by the need to detect galaxies with luminosities down to $\sim 10^5\,L_\odot$. The next leap is expected with the Large Synoptic Survey Telescope \citep[LSST; aka the Vera Rubin Observatory;][]{ive08}. By the end of the decade, LSST should provide a complete sample for distances up to $\sim 1\,\mathrm{Mpc}$ and luminosities down to $\sim 2\times 10^{3}\,L_\odot$, and could detect any novae and supernovae in faint dwarf galaxies out to much greater volumes \citep{2015ApJ...805L...2C}.  The spectroscopic characterization of these satellites will remain a major challenge, probably requiring $30\,$m class telescopes, and no survey will be able to observe within $\sim \pm 10^\circ$ of the galactic plane~\citep{Joshua:2019}.

The Laser Interferometer Space Antenna \citep[LISA;][]{LISA_ESA:2017} is a millihertz gravitational-wave (GW) observatory planned for launch in 2034. LISA will survey the entire sky with a depth of a few hundred kpc for double white dwarfs (DWDs) and other solar-mass binary compact objects with orbital periods $\lesssim 10$~min~\citep{KorolEtAl:2018}. 

In this Letter we show that LISA could provide new and complementary information about MW satellites using  populations of short-period DWDs as tracers of these dwarf galaxies, and as markers of the astrophysical processes and conditions within their unusual (compared to the MW) environments. We will also show that LISA will observe tens of DWDs within the Large and Small Magellanic Clouds (LMC and SMC) and will unambiguously place them within specific regions of the Clouds.  A handful of short-period DWDs should also be observable in other satellites. If located above $\sim 30^\circ$ of the galactic plane, they can be easily associated to their host, since galactic DWD foreground sources are rare. At frequencies above a few mHz, LISA can also probe the zone of avoidance around the galactic plane.

\Needspace{3\baselineskip}
\section{Expected DWD population}

%
%
\begin{table*}[tbh!]
\centering
\caption{Promising satellites for GW detection. Mass, distance, and sky location are taken from \citet{mcconnachie12,LMCSMC,Joshua:2019}. The expected number of LISA sources is estimated using the models of \cite{SatPopAstroPaper}. The sky localization is the $90\%$ area recovered for the fiducial DWD described in \autoref{sec:sensitivity} for each host satellite. We assume a 4 year mission duration and the SciRD noise spectral density \citep{LISAscireq}.
}
\hspace{-1.2cm}
\begin{tabular}{l c c c c c}
\toprule
& LMC & SMC & Sagittarius & Fornax & Sculptor  \\
\midrule
Stellar Mass $(M_\odot)$ & $1.5 \times 10^9$ & $4.6 \times 10^8$ & $2.1 \times 10^7$ & $2.0 \times 10^7$ & $2.3 \times 10^6$  \\
Distance (kpc) & 50.0 & 60.6 & 26.7 & 139 & 86  \\
Ecliptic latitude $\beta$  & -85.4$^\circ$ & -64.6$^\circ$ & -7.6$^\circ$ & -46.9$^\circ$ &-36.5$^\circ$ \\
Galactic latitude $b$  & -32.9$^\circ$ & -44.3$^\circ$ & -14.2$^\circ$ & -65.7$^\circ$ & -83.2$^\circ$ \\ 
Galaxy area (deg$^2$) & 77 & 13 & 37 & 0.17 & 0.076  \\
\midrule
Foreground sources & 1 & 0.2 & 20 & $10^{-3}$ & $3\times10^{-4}$  \\
Expected sources (optimistic) & $>$100 & $>$25 & 10 & 0.2 & 0.07  \\
Expected sources (pessimistic) & 70 & 15 & 3 & 0.1 & $<$0.04  \\
Sky localization (deg$^2$) & 2.1 & 3.1 & 2.3 & -- & 9.3  \\
\bottomrule
\end{tabular}
\label{table:known_satellites}
\end{table*}
%
%

To date no undisputed DWD is known in MW satellites. An X-ray source, RX J0439.8-6809, has been tentatively identified as a compact accreting WD system with a He WD donor in the LMC~\citep{1994A&A...281L..61G,1997A&A...323L..41V}, although later spectral modeling suggests this object may also be consistent with an unusually hot WD in the MW halo~\citep{2015A&A...584A..19W}.  This lack of observational evidence is due to the faintness of these systems. They are undetectable by optical telescopes at the distance at which satellites are typically found---the median distance of known satellites is $\sim 85$ kpc, see \cite{Joshua:2019}.

\subsection{Astrophysical modeling}

A companion paper by \citet{SatPopAstroPaper} investigates the population of DWDs radiating in the LISA sensitivity band in MW satellite galaxies (see also \citealt{Lamberts19}). In that paper, a suite of models that span metallicity, star formation history (SFH) and unstable mass transfer phase are constructed using the population synthesis code \texttt{SeBa} and calibrated against state-of-the-art observations of DWDs \citep{Por96, Nel01,Too12,Too17}. Here we summarize the main assumptions and results, and we refer the reader to the companion paper for details.

Despite the many uncertainties surrounding the composition and formation history of these satellites, the parameters crucial for determining the number of sources detectable by LISA are: (i) the total stellar mass $M_\star$, which sets the fuel supply used to generate stars and (ii) the star formation history (SFH), which controls the mass and frequency distribution of DWDs within the LISA sensitivity band at the present time.

Star formation histories in MW dwarf satellites vary greatly, ranging from purely old populations (formed over 12~Gyr ago) to constantly star forming \citep[e.g.,][]{bro14,wei14,wei19}. To cover the range of possible SFHs we consider a constant star formation rate of 1\,M$_\odot$\,yr$^{-1}$ and an exponentially decaying one with characteristic timescale $\tau_\mathrm{SF} = 5\,$Gyr \citep{wei14}, as optimistic and pessimistic star formation models, respectively.

By setting the metallicity to $Z=0.01$, the binary fraction to 50\% and the initial mass function to \citet{Kro93}, the optimistic (pessimistic) SFH model predicts 0.2 (0.1) detectable sources for a satellite with $M_\star = 10^7$M$_\odot$ at the distance of 100~kpc. Results scales linearly with the mass of the satellite. Other unconstrained parameters, such as metallicity, binary fraction and unstable mass transfer have very minor impacts on the detectable DWD rate and, together, affect predictions by only a factor of a few. 

\Needspace{4\baselineskip}
\subsection{Known satellites}

\autoref{table:known_satellites} summarizes the properties of selected known MW satellites and the expected number of DWDs that can be observed by LISA according to the population synthesis models. We assume a mission duration $T_\mathrm{obs} = 4\,\mathrm{yr}$ and a noise spectral density corresponding to the LISA Science Requirements Document \citep[SciRD;][]{LISAscireq}. The choice of noise spectral density has a significant effect on the number of sources expected---the SciRD sensitivity curve is a factor 1.15, 1.4, and 1.5 \emph{worse} than the original LISA noise curve \cite[see Figure 1 in][]{LISA_ESA:2017} at 3, 5, and $\gtrsim$~10 mHz, respectively. Using the more optimistic noise curve of \cite{LISA_ESA:2017}, as in many previous studies \citep[e.g.][]{bur19,kor17,rob17},  would roughly double the expected number of sources.

The number of DWDs that we expect to see in a particular satellite depends strongly on the mass of the satellite, on its SFH, and on its distance. It depends somewhat less strongly on the ecliptic latitude of the satellite via the weakly directional ``pointing'' of the LISA instrument. The  Magellanic Clouds and the Sagittarius, Fornax, and Sculptor dwarf spheroidal galaxies are promising systems to host detectable LISA sources \citep{SatPopAstroPaper}.

The LMC and SMC are by far the most massive known satellites of the MW.  They are expected to contain $10^2\textrm{--}10^3$ detectable DWDs~\citep{SatPopAstroPaper}. Sagittarius is expected to host several detectable sources, even for a pessimistic SFH model. The rates for Fornax, Sculptor, and smaller galaxies are lower, but these predictions depend on the specific details of the SFH.  

Other satellites can be reached by LISA, but may already have exhausted their reservoir of observable DWDs. LISA is thus in a position to study details of the LMC/SMC, detect a handful of DWDs in some of the more massive satellites, and identify systems in other satellites if they have undergone recent star formation. Furthermore, LISA has the unique opportunity to discover new MW satellites. 

%
%
\begin{table*}[tbh!]
\centering
\caption{Parameters used in our 4200 runs. We grid over these parameters as well as our sample of 56 dwarf galaxies. }
\label{table:run_params}
\hspace*{-1cm} 
\begin{tabular}{l c c c}
(a) Masses \\
\toprule
  & $m_1 \, (M_\odot)$ & $m_2 \, (M_\odot)$ & $\mathcal{M} \, (M_\odot)$    \\
\midrule
He WDs & 0.4 & 0.35 & 0.33 \\
Typical WDs & 0.6 & 0.55 & 0.5 \\
Heavier WDs & 0.7 & 0.65 & 0.59 \\
Extremely low-mass & 0.7 & 0.2 & 0.31 \\
Type Ia SN progenitors & 0.9 & 0.85 & 0.79 \\
\bottomrule
\end{tabular}
\begin{tabular}{c}
(b) Frequencies \\
\toprule
 $f_0$ (mHz)   \\
\midrule
2 \\ 3 \\ 4 \\ 5 \\ 10 \\
\bottomrule
\end{tabular}
\begin{tabular}{l c}
(c) Inclinations \\
\toprule
  & $\iota$ (rad)  \\
\midrule
face-on & 0  \\
intermediate & $\pi/3$  \\
edge-on & $\pi/2$  \\ \\ \\
\bottomrule
\end{tabular}
\end{table*}
%
%

\section{LISA signal recovery}

Having established that LISA can and will observe DWDs hosted by MW satellite galaxies, we need to consider whether it will be possible to associate these DWDs with the actual host satellite.  The challenge is further exacerbated by the fact that LISA will observe ten to fifty thousand galactic DWDs, in addition to the unresolved stochastic foreground produced by $\sim 10^6$ DWDs. Making these associations will depend on LISA's ability to measure source sky locations and distances for these sources. We investigate this by performing full parameter estimation analyses on mock LISA data which we generate for a range of plausible sources within the satellites.

We consider a number of DWD systems that we expect to populate these satellites, spanning mass, frequency and binary inclination (see \autoref{table:run_params}). For each choice of these parameters (75 combinations in total) we place a binary randomly within each of the 54 satellites in \citet{Joshua:2019}, together with the LMC and SMC. The distance and angular size of these satellites are taken from \citet{mcconnachie12} and \citet{LMCSMC}. 

We generate mock LISA data sets lasting ${T_\mathrm{obs} = 4\,\mathrm{yr}}$ and containing the individual DWDs with zero noise. We recover the sources using the conservative LISA SciRD noise power spectral density \citep{LISAscireq} generated with LISACode \citep{lisacode}, with an estimation of the galactic confusion noise taken from~\cite{Babak:2017tow}. 

Gravitational radiation from the DWDs is treated as a quasi-monochromatic signal with linear drifts in frequency:
\begin{equation}
f_\textsc{GW}(t) = f_0 + \dot{f}_0 (t - t_0). 
\end{equation}
We model the effect of these GW signals on the three noise-orthogonal channels $A$, $E$ and $T$~\citep{PhysRevD.66.122002,2014LRR....17....6T}, and process the resulting data using a coherent Bayesian analysis. Each of the signals is described by 8 unknown parameters:
\begin{equation}
    \{\mathcal{A}, f_0, \dot{f}_0, \lambda, \beta, \iota, \psi, \phi_0\}\,, \label{eq:params}
\end{equation}
where $\mathcal{A}$ is the GW amplitude, $(\lambda, \beta)$ are the ecliptic longitude and latitude, respectively, $\iota$ is the inclination angle, $\psi$ is the polarization angle, and $\phi_0$ is an arbitrary initial phase.  The GW amplitude is given by
\begin{equation}
    \mathcal{A} = \frac{2 (G \mathcal{M})^{5/3} }{c^4 D} (\pi f_0)^{2/3}.
\end{equation}
This is set by the source's distance $D$ and chirp mass
\begin{equation}
    \mathcal{M} = \frac{(m_1 m_2)^{3/5}}{(m_1 + m_2)^{1/5}}, 
\end{equation}
for component masses $m_1$ and $m_2$.

For each signal injection, the GW amplitude, frequency, sky position, and inclination are chosen from our grid defined in~\autoref{table:run_params}.  The polarization and initial phase are chosen randomly with a flat distribution.  Finally, $\dot{f}_0$ is chosen according to the gravitational radiation reaction:
\begin{equation}
\label{fdot}
    \dot{f}_0 = \frac{96}{5}\frac{(G\mathcal{M})^{5/3}}{\pi c^5} (\pi f_0)^{11/3}.
\end{equation}

During parameter estimation, we treat $\dot{f}_0$ as an unknown parameter which can take either positive or negative values to account for the possibility of accretion affecting the period evolution of the system \citep{2017ApJ...846...95K}. Population synthesis studies predict that $< 10 \%$ of DWD systems in this frequency range will be in mass-transfer states \citep[e.g.,][]{nel04}.  Priors are chosen to be flat in $\log \mathcal{A}$, $\sin \beta$, $\cos \iota$, and flat in all other parameters in \autoref{eq:params}. 

Our grid over parameters and satellites covers a range of sources from the very quiet to the very loud.
We consider a source to be detected if the coherent signal-to-noise ratio exceeds 7.  This is a conservative choice---\cite{CrowderCornish} and \cite{PhysRevD.81.063008} propose detection thresholds of 5 and 5.7, respectively, for monochromatic sources.  Of our 4200 injected sources, 1954 are detected. For all satellites, at least one combination of the parameters produces a detectable DWD. For nearby satellites, a large range of parameters produce detectable systems. 

%
%
\begin{figure*}
    \centering
    \includegraphics[width=\linewidth]{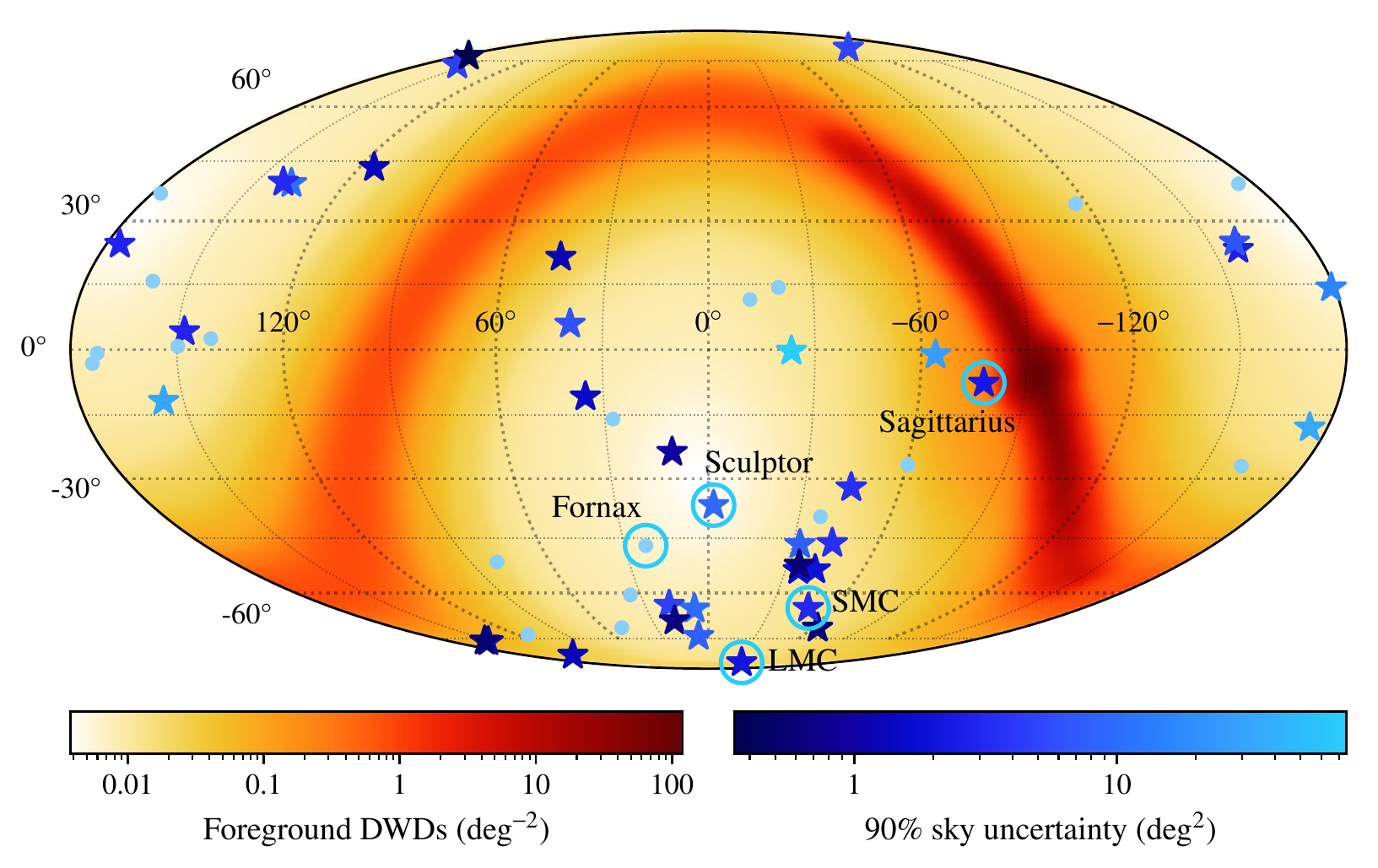}
    \caption{LISA sensitivity to  a fiducial source with $\mathcal{M} = 0.5 \, M_\odot$, $f = 5$~mHz, and $\iota = \pi/3$ in each satellite. Light blue dots are `undetected' sources ($\rho < 7$). Stars are `detected' sources ($\rho > 7$), and are color-coded according to the quality of their sky localizations. For a fixed frequency, this is largely governed by distance and ecliptic latitude. The largest sky uncertainty, for a source with $\rho=7.6$, $D=118$~kpc, and $\beta=-0.3^\circ$, is 75~deg$^2$.  The smallest, for a source with $\rho=36$, $D=22$~kpc, and $\beta=77^\circ$, is 0.3~deg$^2$. Satellites of interest are highlighted with blue circles. In this case, LISA is sensitive to systems at distances of $\lesssim 120$~kpc, which excludes Fornax. } 
    \label{fig:ang_res_recovered}
\end{figure*}
%
%

To help summarize our results, hereafter we will focus on the following five satellites: the LMC, SMC, Sagittarius, Sculptor, and Fornax (see \autoref{table:known_satellites}). 
These satellites span a broad range in distance, ecliptic latitude, and angular scale and are the most likely to host detectable DWDs \citep{SatPopAstroPaper}.
We will also focus our discussion on a fiducial source with a chirp mass of $\mathcal{M}=0.5 \, M_\odot$, radiating at $f_0=5$~mHz, and with an intermediate inclination of $\iota=\pi/3$.  
The detectability of such a source for the complete set of satellites is shown in \autoref{fig:ang_res_recovered}. 
If any such system is present within $\sim$120~kpc, it will be detectable by LISA. 
This represents roughly half the known MW satellites, including all our highlighted satellites except Fornax, which is at a distance of 139~kpc.

\section{Host satellite identification}
\label{sec:sensitivity}

We have shown that LISA will be sensitive to DWDs radiating at a few mHz in the MW satellites. However, it is not immediately obvious that these sources can be robustly associated with their host satellites. In this section, we will consider three pieces of information to solve this problem: the source sky localization, the anisotropic distribution of foreground MW DWDs, and measurements of the source distance. 

At mHz frequencies, LISA's angular resolution is good.  Most injections of our fiducial source can be located to within $\lesssim$10~deg$^2$ (see \autoref{fig:ang_res_recovered}); the exceptions are low-SNR sources near the ecliptic.  This means that sources inside the LMC, SMC, and Sagittarius (all of which are larger than 10~deg$^2$), can potentially be localized to specific regions of the satellites.  The sky uncertainty depends strongly on the SNR and GW frequency ($\propto \rho^{-2} f_0^{-2}$), and also on the ecliptic latitude (a source on the ecliptic has an order of magnitude more uncertainty than a source at the poles).  

Equally important to the sky localization is the foreground of MW DWDs for each satellite.  At frequencies $\gtrsim 3$~mHz, MW DWDs become resolvable \citep{Babak:2017tow}, so the stochastic MW foreground is not a significant concern here. We model the MW sources following \citet{kor19}, including a stellar halo generated with a single burst SFH, a power law density distribution according to \citet{ior18}, and a total mass of $1.4 \times 10^9\,$M$_\odot$ \citep{mac19}. The resulting foreground is strongly anisotropic, closely following the galactic plane (see \autoref{fig:ang_res_recovered}). Most known satellites are well away from the galactic plane, in regions with a foreground density of $\sim 0.01/\text{deg}^{2}$. For a sky localization of $\sim 1$--$10$~deg$^2$, this corresponds to $\sim 0.01$--$0.1$ contaminating foreground sources, or a typical false alarm probability between $\sim 5\times 10^{-5}$ and $\sim 5\times 10^{-3}$. At lower (higher) frequencies, the sky localization is worse (better) and the false alarm rate rises (falls).  

\begin{figure*}
    \centering
    \includegraphics[]{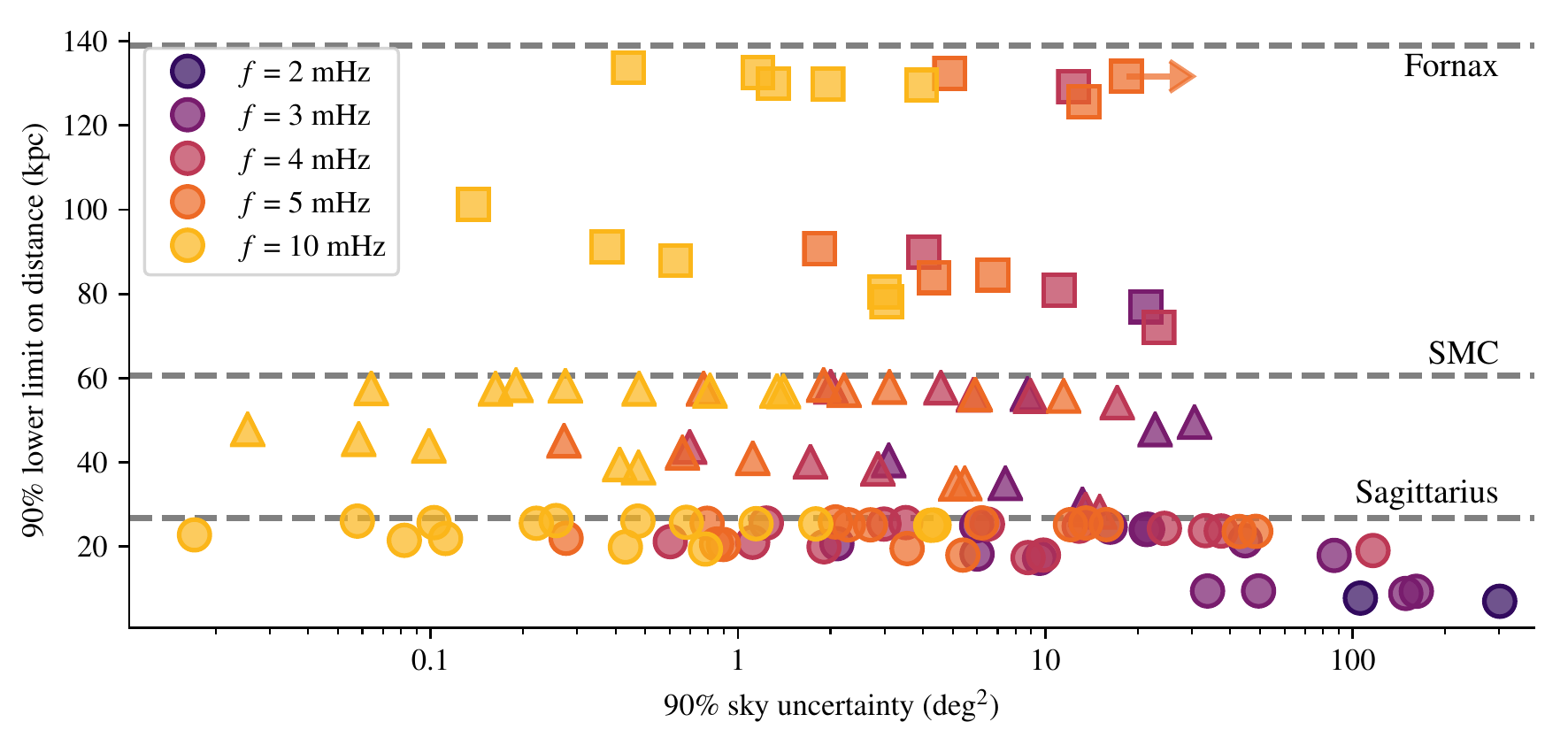}
    \caption{Distance lower limits and sky localizations for all `detected' runs in Sagittarius (circles), the SMC (triangles), and Fornax (squares).  Dashed lines mark the true distance of each satellite. For a given frequency, the sky localization is primarily affected by the source's mass, and the lower limit on the distance is primarily affected by the source's inclination. Lower limits on the sky localization are given for one source which does not have a well-defined 90\% sky area, but does have a well-defined area at lower confidence. }
    \label{fig:dist_area}
\end{figure*}

In addition to associations based on the sky localization, the frequency evolution for sources above $3$--$4$~mHz will be measurable. Our 90\% fractional errors on $\dot{f}$ are distributed according to:
\begin{equation}
    \Sigma_{\dot{f}} \approx 0.07\, \left(\frac{\rho}{10}\right)^{-1} \left(\frac{f}{5\text{ mHz}}\right)^{-11/3} \left(\frac{\mathcal{M}}{ 0.5 \, M_\odot}\right)^{-5/3} \!\!.
\end{equation}
Assuming the inspiral is driven by radiation reaction (\autoref{fdot}), measurements of $\dot{f}$ and $\mathcal{A}$ permit the measurement of the distance to the satellite with a precision of $\sim 30\%$. Stellar interactions within DWDs will reduce, not increase, the total $\dot{f}$ \citep{ 2017ApJ...846...95K}. This implies that a lower limit can safely be set on the distance (see \autoref{fig:dist_area}), thereby further reducing the chance of a false positive. 

Let us examine some cases in detail.  The LMC and SMC are large satellites with many expected sources. Both galaxies have large angular extents and high ecliptic latitudes, so sub-galaxy localizations of sources are likely.  The SMC is in a region of the sky with very few foreground sources, so statistical associations can be readily made.  This is also true for the LMC, but its situation is complicated by a partial degeneracy in the LISA response at extreme ecliptic latitudes. This may result in a larger foreground than stated in \autoref{table:known_satellites}, due to the presence of MW DWD sources at the other ecliptic pole.  Distances for sources in the Magellanic clouds should be well-measured, and will help make associations robust.

Sagittarius is a relatively massive and nearby dwarf spheroidal galaxy, so several detectable sources are expected. Its unusual SFH means that the `optimistic' case of 10 sources from \autoref{table:known_satellites} is quite plausible \citep[see Section 3.3 of][]{SatPopAstroPaper}.  Unfortunately, its location near the galactic bulge and its large angular scale lead to a large number of foreground sources (although note that the foreground varies by an order of magnitude across the satellite).  Frequency measurements can be used to partially remove the foreground. If we consider only sources with $f > 3$~mHz, the foreground drops from 20 to 5.  Distances for this satellite are likely to be well-measured, but as Sagittarius is well within the MW halo, their additional constraining power will be somewhat reduced. Robust associations with Sagittarius will be non-trivial, but careful modeling of the MW population should make it possible.

Fornax, Sculptor, and the other dwarf satellites are too small or too distant to be likely hosts of LISA sources. However, it is possible that uncertainties in the SFH, perhaps combined with a more optimistic LISA noise curve will produce detectable DWDs. In this case, the sources should be readily identifiable. The foregrounds are small, and distance lower limits (particularly in the case of Fornax) would provide strong evidence for the satellite association.

\section{Discovering hidden satellites of the Milky Way}

Unlike light, GWs are not impeded by dust and gas. Moreover, above a few mHz, DWDs become individually resolvable and the MW no longer acts as a GW confusion-noise foreground. This gives LISA an advantage over EM telescopes in that it can peer through the galactic plane and possibly make discoveries on the far side. Currently, the best example of a satellite near the galactic plane is the recently discovered Antlia 2 which has a galactic latitude of $\sim 11^\circ$~\citep{Antlia2-discovery}.  However, at lower latitudes dust extinction increases dramatically; therefore, even objects as large as the LMC could have remained undetected.

If such an object exists, LISA could potentially detect high-frequency DWDs from it. The question is then whether these detections are sufficient to infer the presence of the hidden satellite. This task is complicated by the high density of resolvable foreground DWD sources in the galactic plane. 

To illustrate the discovery potential of LISA, consider a hypothetical satellite, similar to the LMC, at a distance of $50$~kpc behind the disk of the MW. We assume that it has an angular diameter of $10^\circ$, a mass of $1.5\times 10^{9}\,M_{\odot}$, a fixed metallicity of $Z=0.005$, a constant star formation rate, and an age of 13.5~Gyr  \citep[c.f.][]{SatPopAstroPaper}. This object could be completely covered by the galactic disk.

The foreground density of DWD sources in the disk is $\sim 100/ \text{deg}^2$ (see \autoref{fig:ang_res_recovered}).  If galactic sources are distributed uniformly throughout the disk, which has a total area of $\sim 3000\text{ deg}^{2}$, then a simple Poisson counting argument suggests that an excess of $\sim 100$ sources in an $80\text{ deg}^2$ patch of the sky would be a significant overdensity at the $90\%$ level. 

We expect $\sim 100$ sources in our hypothetical satellite, so an LMC-like satellite at $\lesssim 50$~kpc should appear as a statistically significant overdensity.  At greater distances, it would have too few sources to overcome the foreground. This calculation assumes a similar stellar density to the LMC; a denser (sparser) satellite would be detectable at a greater (lower) maximum distance. Furthermore, we assume a uniform, Poissonian distribution of DWDs in the galactic disk---a more realistic non-uniform distribution will require a larger overdensity to be significant. 

However, we have not yet considered distance measurements. \autoref{sec:sensitivity} suggests that the majority of detectable extragalactic sources will be chirping, meaning that lower bounds can be placed on their distances. This will allow us to distinguish them from the foreground and detect satellites out to greater distances. We assume that chirping sources allow us to place a lower limit on the distance of $\sim60\%$ of the true value (although many sources do considerably better---see \autoref{fig:dist_area}).  A satellite at 50~kpc with multiple detected sources can be confidently placed at $\gtrsim 30$~kpc, which is greater than the distance to any DWD in the galactic disk.  At 150~kpc (200~kpc) we expect to detect $\sim 10$ ($\sim 3$) sources from our hypothetical satellite which can likewise be distinguished from disk foreground sources (although a small number of halo sources remain as contaminants).

The galactic plane obscures $\sim 10\%$ of the sky. For the first time, LISA will be able to survey this region for major MW satellites out to astrophysically interesting distances of $\lesssim 200$~kpc.

\section{Conclusions}

We have shown that if a population of DWDs emitting GWs at $\gtrsim 3$~mHz exists in the MW satellites, LISA will be able to detect them. Although the exact rate depends on the star formation history of each satellite, it is probable that many such DWDs will be detected in several different satellites.  Moreover, in this frequency band, LISA will provide sky localizations of $\sim10 \text{ deg}^2$ and distance measurements with errors of $\sim 30\%$.  This means that LISA should be able to associate these DWDs to their host satellites. Finally, at frequencies above a few mHz, the galactic confusion noise clears, and LISA can see through the galactic disk and bulge.  This fact, combined with the arguments above, suggests that LISA might be capable of discovering hidden satellites of the MW, provided they are sufficiently massive.

Observations of short-period extragalactic DWDs will naturally occur as part of the LISA survey of the galactic DWD population.  These observations will complement those of large optical surveys, since the selection effects are very different. The possibility of detecting short-period DWDs in MW satellites highlights the discovery space opened up by a GW observatory and its potential impact on a wide range of open questions in astrophysics and cosmology, from low-metallicity star formation history and heavy element nucleosynthesis to small-scale cosmology in the nearby Universe.

\vspace{-10pt}

\acknowledgments{
We thank Christopher Berry, Siyuan Chen, Sean McGee, Hannah Middleton, Patricia Schmidt, and Alberto Sesana for discussions. DG is supported by Leverhulme Trust Grant No. RPG-2019-350.  VK and ST acknowledge support from the Netherlands Research Council NWO (respectively Rubicon 019.183EN.015 and VENI 639.041.645 grants). TEW acknowledges support from the NRC-Canada Plaskett fellowship. AV acknowledges support from the UK Space Agency, the Royal Society and the Wolfson Foundation. Computational work was performed on the University of Birmingham BlueBEAR cluster, the Athena cluster at HPC Midlands+ funded by EPSRC Grant No. EP/P020232/1, and the Maryland Advanced Research Computing Center (MARCC).
}
\vspace{-10pt}

\pagebreak

\software{We make use of \texttt{LISACode} \citep{lisacode} and the LISA Data Challenge software (\url{https://lisa-ldc.lal.in2p3.fr/}), along with the \texttt{python} packages \texttt{astropy} \citep{astropy:2013,astropy:2018}, \texttt{healpy} \citep{healpix}, \texttt{cpnest} \citep{cpnest}, and \texttt{ligo.skymap} \citep{2016ApJS..226...10S}.  Posteriors and other data products for this work will be made available at \url{https://doi.org/10.5281/zenodo.3668905}.}

\vfill
\null

\bibliography{references}

\end{document}